\documentclass[12pt]{iopart}

\usepackage{iopams}  

\def\beq{\begin{equation}}
\def\eeq{\end{equation}}
\def\bea{\begin{eqnarray}}
\def\eea{\end{eqnarray}}
\def\nn{\nonumber}
\def\g#1#2{{\mathfrak g}_{#1}(#2)}
\def\hf{\frac{1}{2}}
\def\g#1#2{{\mathfrak g}_{#1}(#2)}
\def\hg#1#2{\hat{\mathfrak g}_{#1}(#2)}
\def\cg{ \check{\mathfrak g}_{\ell} }
\def\N{ {\mathbb N} }
\def\Pop#1#2{P^{(#1)}_{#2}}
\def\PM#1#2{\hat{P}^{(#1)}_{#2}}
\def\PE#1#2{\check{P}^{(#1)}_{#2}}
\def\Pone#1{P^{(#1)}}
\def\ket#1{\left| #1\right\rangle}
\def\v0{ \ket{\delta,r,\mu} }
\def\u0{ \ket{\delta,r,\theta} }
 
\def\a{\underline{a}}
\def\b{\underline{b}}
\def\ep#1{\underline{\epsilon}_{\,#1}}
\def\de#1{\underline{\delta}_{\,#1}}
\def\der#1{\frac{\partial}{\partial #1} }
\def\del#1{\frac{\partial}{\partial #1}}
\def\dels#1{ \frac{\partial^2}{\partial #1^2} }
\def\binom#1#2{ \left(\begin{array}{c} #1 \\ #2 \end{array}\right) }

\begin{document}

\title[$\ell$-Galilei invariant equations]{Intertwining operators for $\ell$-conformal Galilei algebras and hierarchy of invariant equations
}

\author{N Aizawa$^1$, Y Kimura$^1$, J Segar$^2$}

\address{$^1$  
Department of Mathematics and Information Sciences, 
Graduate School of Science, 
Osaka Prefecture University, 
Nakamozu Campus, Sakai, Osaka 599-8531, Japan}

\address{$^2$ 
Department of Physics, 
Ramakrishna Mission Vivekananda College, 
Mylapore, Chennai 600 004, India}

\eads{\mailto{aizawa@mi.s.osakafu-u.ac.jp}, \mailto{segar@rkmvc.ac.in}}

\begin{abstract}
 $ \ell$-Conformal Galilei algebra, denoted by $ \g{\ell}{d}, $ is a non-semisimple Lie algebra 
specified by a pair of parameters $ (d,\ell).$ The algebra is regarded as a nonrelativistic analogue of 
the conformal algebra. We derive hierarchies of partial differential equations which have invariance of the group 
generated by $ \g{\ell}{d} $ with central extension as kinematical symmetry. 
This is done by developing a representation theory such as Verma modules, singular vectors of $ \g{\ell}{d} $ and 
vector field representations for $ d = 1, 2.$ 
\end{abstract}

\submitto{\JPA}

\section{Introduction}

  When a differential equation is given, searching for its symmetry is 
one of the standard approach to find solutions to the equation or to 
study its properties. 
Sometimes it turns out that the given equation has larger symmetry than 
expected from physical intuition. Maximal kinematical symmetry of 
Schr\"odinger equation for a free particle is one of such cases. 
It is obvious that Galilei group is a symmetry of the free Schr\"odinger equation. 
However, Niederer showed that the equation is also invariant under the scale  and 
the special conformal transformations \cite{Nie} (see also \cite{Ha}). 
The Lie group of maximal kinematical symmetry of free Schr\"odinger equation is nowadays 
called Schr\"odinger group. 
Although the Schr\"odinger group has been known since the period of  Lie, 
it is not so long time ago that the group is recognized as one of the fundamental algebraic object in physics. 
The group and its Lie algebra play certain roles in wide range of 
physical problems from fluid dynamics to nonrelativistic holography. 
One may find a very nice review with an appropriate list of references on 
physical implications of Schr\"odinger group and its Lie algebra in \cite{SV} (see a Forward by Henkel). 

  Returning to the problem of symmetry of a given differential equation, 
one may reverse the problem. 
Namely, for a given Lie group we look for differential equations invariant under the group. 
When the invariance means usual \textit{Lie symmetry} one may use the algorithm based on 
the classical Lie method (see e.g. \cite{Ovs,Olv}). 
Especially wide class of equations invariant under the Schr\"odinger group has been obtained by the algorithm \cite{FuCh0,RiWin,FuChe2}. 

%

 In the present work, we consider $\ell$-conformal Galilei group as a kinematical symmetry for a hierarchy of differential 
equations. It is defined as the symmetry under the transformation of the independent variables $ x_i \to \tilde{x}_i $ together with 
the transformation of a solution $ \psi(x_i) $ to an another solution of the form $ f(\tilde{x}_i) \psi(\tilde{x}_i) $ where 
$ f(x_i) $ is a weight function independent of $ \psi $ \cite{Nie}. 
As we shall see such equations are obtained as a byproduct of the representation theory of Lie algebra of the group. 
In fact, this is an established fact if the given Lie group is real connected and semisimple. \cite{Kos,Dob}. 
Although the Lie algebra of  Schr\"odinger group is not semisimple, it is known that 
similar techniques to semisimple case is applicable to develop its 
representation theory \cite{DoDoMr, Mur}.  
This enable us to apply the method similar to \cite{Kos,Dob} to find differential equations having the 
Schr\"odinger group as kinematical symmetry  
\cite{DoDoMr,Mur,ADD,ADDS}. 
Here we extend this analysis to larger group,  $\ell$-conformal Galilei group with central extension. 
We develop representation theory of th Lie algebra of the $\ell$-conformal Galilei group  
and, as a consequence, derive differential equations having the group as kinematical symmetry. 

By definition Schr\"odinger group is an enlargement of Galilei group. 
It has been shown that one can further enlarge Schr\"odinger group 
and has a sequence of Lie groups in which each Lie group is specified by a 
parameter $ \ell \in \hf {\mathbb N} $, where ${\mathbb N}$ denotes the set of positive integers \cite{HaPle,NdelOR}. 
We refer to each group of the sequence as $ \ell$-conformal Galilei group. 
Here we assumed that the dimension of spacetime is fixed. 
Taking into account the difference of spacetime dimension one may say that each $ \ell$-conformal Galilei group 
is specified by a pair of parameters $ (d,\ell) $ where dimension of spacetime is $ (d+1).$ 
The first member with $ \ell = 1/2$ of the sequence corresponds to the Schr\"odinger group. 
As the Schr\"odinger group, $ \ell = 1 $ conformal Galilei group and its Lie algebra also appear in wide range of physical problems. 
For instance, nonrelativistic  electrodynamics \cite{NdORM2}, classical mechanics with 
higher order time derivatives \cite{StZak,LSZ,LSZ2}, 
nonrelativistic analogue of AdS/CFT correspondence \cite{MT,BG,ADV,BaMa,BaKu}, 
nonrelativistic spacetime and gravity \cite{DuHo,HoKuNi,Bagchi,BaFare,BaGoGo}, 
quantum mechanical particle systems \cite{AGKP}, conformal mechanics \cite{FeIvLu,GaMa2013}, 
non-relativistic twistors \cite{FeKoLuMas} and so on. 
Furthermore, one may find applications of the algebraic structure to mathematical
studies of topics such as systems of partial differential equation \cite{FuChe2,FuChe,ChHe}. 
It is, however, quite recent that conformal Galilei groups and algebras with higher $ \ell $ are 
studied from physical and mathematical points of view \cite{GalMas2011,DH,GomKam2012,AnGoMas,AIK,GaMa2013,AnGo2013,AnGoKD}. 
It has been observed that physical systems having a connection with $\ell$-conformal Galilei $(\ell > 1/2)$ group are described by 
Lagrangians or Hamiltonians with higher order derivatives. 

 This paper is organized as follows. In the next section we give a short review of generators of 
$ \ell$-conformal Galilei group and its central extensions. Then we consider the case of $ d= 1 $ and half-integer $\ell.$ 
Representation theory of conformal Galilei algebra for this case has been developed in \cite{AIK} and the necessary items 
to find differential equations with desired symmetry, Verma module and singular vectors in it, have also been obtained. 
With these informations we derive a hierarchy of differential equations having the conformal Galilei group for 
$(d,\ell) = (1, {\mathbb N}+\hf) $ as kinematical symmetry.  
We also present a vector field representation of the conformal Galilei algebra on a coset space. 
In \S \ref{SEC:2mass} and \S \ref{SEC:2exo}, the same analysis is repeated for the case of $ d = 2 $ and arbitrary values of $\ell. $ 
In this case, however, representation theory of the conformal Galilei algebra has not been investigated yet. 
So we shall start with constructing Verma modules and singular vectors. We then derive differential equations with desired 
symmetry for the values of $(2,{}^{\forall}\ell). $ 
A vector field representation of the conformal Galilei algebra is also presented. 
Motivated by the recent active studies of holographic dual of 3D Minkowski spacetime \cite{Bagchi,BaFare} we repeat the same analysis 
for $ (d,\ell)=(1,1), $ the case without central extension, in Appendix.


\section{$\ell$-Conformal Galilei algebra}

 We present $ \ell$-conformal Galilei algebra, denoted by $ \g{\ell}{d}, $  as a set of generators of 
transformation of coordinates in  $ (d+1)$ dimensional spacetime \cite{NdelOR}. 
Consider the transformations generated by 
$$
H = \frac{\partial}{\partial t},\quad 
D=-2t\frac{\partial}{\partial t}-2\ell x_i\frac{\partial}{\partial x_i},\quad 
C = t^2\frac{\partial}{\partial t}+2\ell tx_i\frac{\partial}{\partial x_i},
$$
\beq
M_{ij} = -x_i\frac{\partial}{\partial x_j}+x_j\frac{\partial}{\partial x_i},\quad 
P_i^{(n)} = (-t)^n\frac{\partial}{\partial x_i},
\label{generators} 
\eeq
where $i=1,\ldots,d$ specify the space coordinates and $n=0,\ldots,2\ell$ with 
$ \ell \in \hf {\mathbb N}. $ 
We thus have $ d(d-1)/2 + (2\ell+1)d + 3 $ generators. 
Time translation $H$, space translations $ P_i^{(0)},$ spatial rotations $ M_{ij} $ 
and Galilei transformations $ P_i^{(1)} $ form the Galilei group of nonrelativistic kinematics. 
In addition we have numbers of transformation to a coordinate system with acceleration $ P_i^{(n)} \ (n \geq 2)$ 
according to the values of the parameter $ \ell. $ Scale transformation $D$ and special conformal transformation $C$ 
also depend on $ \ell. $ One may see that space and time are equally scaled only for $ \ell = 1. $ 

 It is immediate to see that the generators (\ref{generators}) have closed commutators and form a Lie algebra. 
Nonvanishing commutators of $ \g{\ell}{d} $ are given by
\begin{eqnarray}
  & & [D, H] = 2H, \qquad [D, C]=-2C, \qquad [C, H]=D,  
  \nonumber \\
  & & [M_{ij}, M_{k\ell} ] = - \delta_{ik} M_{j\ell} - \delta_{j\ell} M_{ik} 
     + \delta_{i\ell} M_{jk} + \delta_{jk} M_{i\ell},
  \nonumber \\
  & & [H, \Pop{n}{i}] = -n \Pop{n-1}{i}, \qquad\qquad 
      [D, \Pop{n}{i}] = 2(\ell-n) \Pop{n}{i}, 
  \label{BosonicDef} \\
  & & [C, \Pop{n}{i}] = (2\ell-n) \Pop{n+1}{i}, \qquad
      [M_{ij}, \Pop{n}{k} ] = -\delta_{ik} \Pop{n}{j} + \delta_{jk} \Pop{n}{i}.
  \nonumber
\end{eqnarray}
The algebra  $ \g{\ell}{d} $ has a subalgebra consisting of a direct sum of 
$ sl(2,{\mathbb R}) \simeq so(2,1) $ generated by $ H, D, C $ and 
$ so(d) $ by $ M_{ij}. $ 
While $ P_i^{(n)} $ forms an Abelian ideal of $ \g{\ell}{d} $ so that $ \g{\ell}{d} $ 
is not a semisimple Lie algebra. 

 It is known that $ \g{\ell}{d} $ has two distinct type of central extensions according 
to the values of $d$ and $\ell:$ 
\\ \noindent
(i) mass extension existing for any $ d $ and  $ \ell \in {\mathbb N} + \hf $
\begin{equation}
 [\PM{m}{i}, \PM{n}{j}] = \delta_{ij} \, \delta_{m+n,2\ell}\,  I_m M, \qquad
 I_m = (-1)^{m+\ell+\hf} (2\ell-m)! \, m!.
 \label{MassExtension}
\end{equation}
(ii) exotic extension existing only for $ d = 2 $ and  $ \ell \in {\mathbb N} $
\begin{equation}
  [\PE{m}{i}, \PE{n}{j}] = \epsilon_{ij} \, \delta_{m+n,2\ell}\,  I_m \Theta, \qquad 
  I_m = (-1)^{m} (2\ell-m)!\,  m!, 
  \label{ExoticExtension}
\end{equation}
where $ \epsilon_{ij} $ is the antisymmetric tensor with $ \epsilon_{12} = 1. $ 
We note that the structure constants are chosen to agree, up to an overall factor, with the ones used in
\cite{GalMas2011} and \cite{GomKam2012}. 
Simple explanation of the existence of two distinct central extensions is found in \cite{MT}. 
The Schr\"odinger algebra considered by Niederer corresponds to $ \g{1/2}{d} $ with the mass central 
extension. The exotic extension was first found for $ \ell = 1 $ in the study of classical mechanics 
having higher order time derivatives \cite{StZak,LSZ,LSZ2}. 

In the following sections we discuss the lowest weight representations (especially Verma modules) of 
$\g{\ell}{d} $ with a central extension. 
Since we are interested in an algebra with central extensions, we denote $ \g{\ell}{d} $ with the 
mass and exotic central extensions by $ \hg{\ell}{d} $ and  $\cg,$ respectively. 
By using the representations we derive differential equations having the group generated by 
$ \hg{\ell}{d} $ or  $\cg$ as kinematical symmetry.


\section{$d=1$ Conformal Galilei group with mass central extension}
\label{SEC:1mass}

\subsection{Equations and symmetry}
\label{EqSymm}

 The simplest example of $\ell$-conformal Galilei algebra with central extension 
is $ \hg{\ell}{1}, $  namely, $ d = 1 $ algebra with 
mass central extension ($\ell \in \N+\hf$).  
Thus we start our construction of invariant equations with this simplest algebra $ \hg{\ell}{1}. $ 
We also outline the method of \cite{Dob} throughout the construction. 
The method consists of two steps. First step is  an abstract representation theory, \textit{i.e.,} 
a Verma module over $ \hg{\ell}{1} $ and singular vectors in it. 
Second step is a realization of the Verma module on a space of $C^{\infty}$ functions 
with a special property. The space of $C^{\infty}$ functions 
is also  a representation space of the group generated by $ \hg{\ell}{1}. $

 Verma modules over $ \hg{\ell}{1} $ and its singular vectors have already been studied in detail \cite{AIK}. 
The fact that one can appropriately define Verma modules over  $ \hg{\ell}{1} $ is, in fact, the reason why 
one may apply the method of \cite{Dob}, which is established for semisimple Lie 
groups, to non-semisimple $ \hg{\ell}{1}. $ 
To complete the first step we summarize the results of \cite{AIK} below 
(Highest weight Verma modules are considered in \cite{AIK}. Here we convert it to 
lowest weight modules for later convenience). 

 We make the following vector space decomposition of $ {\mathfrak g} \equiv \hg{\ell}{1} $ 
\bea
  & & {\mathfrak g}^+ = \langle \; H, \hat{P}^{(0)}, \hat{P}^{(1)}, \cdots, \hat{P}^{(\ell-\hf)} \; \rangle, 
  \nn \\
  & & {\mathfrak g}^0 = \langle \; D, M \; \rangle,  \label{TriDecomp} 
  \\
  & & {\mathfrak g}^- = \langle \; C, \hat{P}^{(\ell+\hf)}, \hat{P}^{(\ell+\frac{3}{2})}, \cdots, \hat{P}^{(2\ell)} \; \rangle.
  \nn
\eea
Here we omit the index for space coordinate. 
Then one may see that $ [{\mathfrak g}^0, {\mathfrak g}^{\pm} ] \subset {\mathfrak g}^{\pm}, $ that is, 
this is an analogue of triangular decomposition of semisimple Lie algebras. 
Suppose that there exists a lowest weight vector defined by
\bea
 & &
  D \ket{\delta,\mu} = -\delta\ket{\delta,\mu}, \quad 
  M \ket{\delta,\mu} = -\mu\ket{\delta,\mu}, \nn \\
 & & 
  X \ket{\delta,\mu} = 0, \quad {}^{\forall} X \in \mathfrak{g}^{-}
  \label{LWV1}
\eea
We define a Verma module $ V^{\delta,\mu} $ over $ \mathfrak g $ by 
\beq 
 V^{\delta,\mu} = \left\{ \left. H^{h}\prod_{j=0}^{\ell-\hf} (\hat{P}^{(\ell-\hf-j)})^{k_j}
 	\ket{\delta,\mu} \ \right| \ h,k_0,k_1,\cdots,k_{\ell-\hf}
   	\in {\mathbb Z}_{\geq 0} \right\}.
 \label{Verma1}
\eeq
The vector space $ V^{\delta,\mu} $ carries a representation of $ \mathfrak g $ specified by $ \delta $ and $ \mu. $ 
In general, the representation is not irreducible (reducible). Reducibility of $ V^{\delta,\mu} $ is detected by singular vectors. 
Singular vector is a vector having the same property as $ \ket{\delta,\mu} $ but different eigenvalues. 
If the Verma module has a singular vector, then the representation is reducible. 
For the Verma module given by (\ref{Verma1}) one may prove the following: 
If $ 2\delta-2(q-1)+(\ell+\hf{1})^2=0 $ for $ q \in {\mathbb Z}_{\geq 0} $ then $ V^{\delta,\mu} $ has precisely one singular 
vector given by
 \beq 
  \ket{v_{q}} = {\cal S}^q\ket{\delta,\mu}, \quad {\cal S} = a_{\ell}\mu H + (\hat{P}^{(\ell-\hf)})^2, \quad 
  a_{\ell}=2 \left( \Big(\ell-\hf\Big)! \right)^2
  \label{sing1}
 \eeq
Namely, $ \ket{v_q} $ satisfies the relations:
\bea
  &  & D \ket{v_{q}} = (2q-\delta) \ket{v_{q}}, \quad M \ket{v_{q}} = -\mu \ket{v_{q}},
  \nn \\
  & & X \ket{v_q} = 0, \quad {}^{\forall} X \in \mathfrak{g}^{-} 
  \label{SV1}
\eea
One may built an another Verma module $ V^{2q-\delta,\mu}$ on the singular vector $ \ket{v_q} $ 
by replacing $ \ket{\delta,\mu} $ with $ \ket{v_q} $ in (\ref{Verma1}). 

  We now proceed to the second step. 
We introduced a vector space decomposition (\ref{TriDecomp}) of $ \mathfrak g. $ 
It follows that the Lie group $G$ generated by $ \mathfrak g $ also has the corresponding decomposition, 
$ G = G^+ G^0 G^- $ where $ G^{\pm} = \exp( {\mathfrak g}^{\pm}) $ and $ G^0 = \exp({\mathfrak g}^0). $ 
Consider a $ C^{\infty} $ function on $ G $ having the property of \textit{right covariance}:
\beq
  f(g g^0 g^-) = e^{\Lambda(X)} f(g), \quad {}^{\forall} g \in G, \ {}^{\forall} g^{0} = e^X \in G^{0}, \ {}^{\forall} g^{-}\in G^{-}
  \label{RightCovariance}
\eeq
where $ \Lambda(X) $ is an eigenvalue of $ X \in {\mathfrak g}^0. $ 
Thus the function $ f(g) $ is actually a function on the coset $ G/G^0G^-. $ 
Now consider the space $ C^{\Lambda} $ of right covariant functions on $G. $ 
We introduce the right action of $ \mathfrak g $ on $ C^{\Lambda} $ by the standard formula:
\beq
  \pi_R(X) f(g) = \left. \frac{d}{d\tau} f(g e^{\tau X}) \right|_{\tau = 0}, 
  \quad 
  X \in {\mathfrak g}, \ g \in G
  \label{rightaction-def}
\eeq
Then it is immediate to verify by the right covariance (\ref{RightCovariance}) that $ f(g) $ has the 
properties of lowest weight vector:
\bea
  & & 
    \pi_R(D) f(g) = \Lambda(D) f(g), \quad \pi_R(M) f(g) = \Lambda(M) f(g), 
    \nn \\
  & & \pi_R(X) f(g) = 0, \quad X \in {\mathfrak g}^-
  \label{Raction1}
\eea
Parameterizing an element of $ G^+ $ as $ g^+ = \exp( tH + \sum_{j=0}^{\ell-\hf} x_j \hat{P}^{(j)}) $ 
the right action of $ H $ and $ \hat{P}^{(k)} $ become differential operators on $ C^{\Lambda}:$
\beq
  \pi_R(H) = \del{t} + \sum_{j=1}^{\ell-\hf} j x_j \del{x_{j-1}}, \qquad 
  \pi_R(\hat{P}^{(k)}) = \del{x_k}.
  \label{Raction1-2}
\eeq
Thus one may construct the Verma module $ V^{\delta,\mu}$ by regarding $ f(g) $  as the lowest weight 
vector $ \ket{\delta,\mu} $ with an identification 
$ \Lambda(D) = -\delta, \ \Lambda(M) = - \mu. $ 
The singular vector $ \ket{v_q} $ corresponds to $ f_q(g) \equiv \pi_R({\cal S}^q) f(g). $ 

  Now we are ready to write down differential equations having the  group $G = \exp(\hat{\mathfrak g}_{\ell}(1)) $ as kinematical symmetry.  
Suppose that the operator $ \pi_R({\cal S}^q) $ has a nontrivial kernel:
\beq
 \pi_R({\cal S}^q) \psi(t,x_i) = \left(
   a_{\ell} \mu \Big( \del{t}+ \sum_{j=1}^{\ell-\hf} j x_j \del{x_{j-1}}\Big) + \frac{\partial^2}{\partial^2 x_{\ell-\hf}}
   \right)^q\psi(t,x_i) = 0
   \label{InvEq1}
\eeq
for some function $ \psi \neq 0. $ Then the differential equations (\ref{InvEq1}) have the desired symmetry. 
This is verified in the following way. 
Consider the left regular representation of $ G $ on $ C^{\Lambda}$ defined by
\beq
 (T^{\delta,\mu}(g) f)(g') = f(g^{-1}g'). \label{LeftRegular}
\eeq
If we take the singular vector $ f_q(g')$ instead of $ f(g') $ in (\ref{LeftRegular}), then 
we have an another representation $ T^{2q-\delta,\mu}. $ 
Furthermore, the operator $ \pi_R({\cal S}^q) $ is an intertwining operator of the representations
\beq
   \pi_R({\cal S}^q)\, T^{\delta,\mu} = T^{2q-\delta,\mu} \, \pi_R({\cal S}^q).
   \label{Intertwiner}
\eeq
If follows that if $ \psi(t,x_i) $ is a solution to the equation (\ref{InvEq1}) then 
a transformed function $ T^{\delta,\mu}(g)\psi $ is also a solution to (\ref{InvEq1}):
\[
  \pi_R({\cal S}^q) (T^{\delta,\mu}(g)\psi) = T^{2q-\delta,\mu}(g) \pi_R({\cal S}^q)  \psi = 0.
\]
Thus the group $G$ is the kinematical symmetry of  equation (\ref{InvEq1}) (see \cite{Dob} for more detail).  
We remark that the symmetry transformations are determined by the left action of $ G $ on $ C^{\Lambda}, $ i.e., Eq.(\ref{LeftRegular}). 
On the other hand the differential equations (\ref{InvEq1}) are obtained by using the right action of $\mathfrak g $ on $ C^{\Lambda},$ 
i.e., Eq.(\ref{Raction1-2}).  Thus (\ref{Raction1-2}) is not the generator of symmetry. 

  We have obtained a hierarchy of differential equations (\ref{InvEq1}). 
For $\ell = 1/2$  (\ref{InvEq1}) recovers a hierarchy of heat/Schr\"odinger equations 
in one space dimension obtained in \cite{DoDoMr,ADD,ADDS}:
\[
  \left( 2\mu \del{t} + \dels{x} \right)^q \psi(t,x) = 0.
\]
The central charge $ \mu $ is interpreted as (imaginary) mass. 
For higher values of $ \ell $ we  observe an interesting deviation from the heat/Schr\"odinger equation and 
the obtained equations are highly nontrivial. 
As an illustration we show the hierarchy of equations for $ \ell = 3/2 $ and $5/2.$ 
\bea
  & &  \left( 2\mu \Big(\del{t} +x_1\del{x_0} \Big) + \dels{x_1} \right)^q \psi(t,x_0,x_1) = 0, 
  \quad \ell = 3/2
  \nn\\
  & & \left( 8\mu \Big(\del{t} +x_1\del{x_0} + 2x_2 \del{x_1} \Big) + \dels{x_2} \right)^q \psi(t,x_0,x_1,x_2) = 0, 
  \quad \ell = 5/2
  \nn
\eea
The first member of the hierarchy $(q=1)$ is always a second order differential equation for any values of $ \ell. $ 
This is a sharp contrast to the results in \cite{GomKam2012,AnGoMas,AnGo2013,AnGoKD} where $ \ell$-conformal Galilei 
algebra is relating to systems with higher order derivatives. 
Other examples of second order differential equations relating to the algebraic structure are found in \cite{MT,ChHe}. 

%
\subsection{Vector field representation of $ \hg{\ell}{1}$}

 We used the right action of $ \mathfrak g $ on the space of right covariant functions to 
derive a hierarchy of differential equations. 
One may, of course, consider the left action of $ \mathfrak g $ on the same space defined by
\beq
 \pi_{L}(X)f(g)= \left. \frac{d}{d \tau}f(e^{-\tau X}g) \right|_{\tau=0},\ \ X\in {\mathfrak g}
 \label{leftaction-def}
\eeq
The left action gives a vector field representation of $ \mathfrak g, $ the algebra with the mass central extension. 
It is highly nontrivial so we give its explicit formula. 
The representation of generators in $ {\mathfrak g}^0 $ is given by
$$
  \pi_{L}(D) = \delta-2t\del{t}-\sum_{j=0}^{\ell-\hf}2(\ell-j)
 	x_{j}\del{x_j}, 
 	\qquad 
 	\pi_{L}(M) = \mu.
$$
Generators in  $ {\mathfrak g}^+ $ are represented as
$$
 \pi_{L}(H) = -\del{t}, \qquad 
 \pi_{L}(\hat{P}^{(k)}) = -\sum_{j=0}^{k}\binom{k}{j}t^{k-j} \del{x_j}, 
$$
and those in  $ {\mathfrak g}^- $ as
\bea
 & & \pi_{L}(C) = t\pi_L(D) + t^2 \del{t} + \frac{\mu}{2} \left( \Big(\ell+\hf \Big)! \right)^2 x_{\ell-\hf}^2  
          -\sum_{j=0}^{\ell-\hf} (2\ell-j) x_j \del{x_{j+1}},
    \nn \\
 & & \pi_{L}(\hat{P}^{(k)}) = \mu  \sum_{j=2\ell-k}^{\ell-\hf}\binom{k}{2\ell-j}
   	I_{2\ell-j}\, t^{k-2\ell+j}x_{j} -\sum_{j=0}^{\ell-\hf}
   	\binom{k}{j}t^{k-j}\del{x_j},
   	\nn
\eea
where $ \binom{k}{n} $ is the binomial coefficient. 
It can be easily verified that these representations satisfy the appropriate commutation relations.

\section{$d=2$ Conformal Galilei group with mass central extension}
\label{SEC:2mass}

\subsection{Verma module and singular vector}

Employing the same procedure as the case of $ \hg{\ell}{1}, $ 
we investigate differential equations having the Lie group generated by 
$ \hg{\ell}{2} $ as kinematical symmetry. 
We remark that $ \ell $ is restricted to half-integer values and we have a 
spatial rotation $ so(2) $ in this case. 
We redefine the generators in order to introduce the decomposition analogous to triangular one:
\beq
  \PM{n}{\pm} = \PM{n}{1} \pm i \PM{n}{2}, \qquad 
  J = - i M_{12}, \qquad M \to 2M.
  \label{NewGen2}
\eeq
Then the nonvanishing commutators of $ \hg{\ell}{2} $ are given by
\bea
  & & [D, H] = 2H, \qquad [D, C]=-2C, \qquad [C, H]=D,  
  \nonumber \\
  & & [H, \PM{n}{\pm}] = -n \PM{n-1}{\pm}, \qquad\qquad [D, \PM{n}{\pm}] = 2(\ell-n) \PM{n}{\pm}, 
  \nonumber \\
  & & [C, \PM{n}{\pm}] = (2\ell-n) \PM{n+1}{\pm}, \qquad \, [J, \PM{n}{\pm}] = \pm \PM{n}{\pm},
  \label{ExoticCGAdef} \\
  & & [\PM{m}{\pm}, \PM{n}{\mp} ] = \delta_{m+n,2\ell}\, I_m M. \nn
\eea
Here $ I_m $ is given by (\ref{MassExtension}). 
We introduce a vector space decomposition of $ {\mathfrak g} = \hg{\ell}{2}:$ 
\bea
  & & {\mathfrak g}^+ = \langle \ H, \ \PM{n}{\pm} \ \rangle, \quad n = 0, 1, \cdots, \ell- \frac{1}{2}
   \nn \\
  & & {\mathfrak g}^0 = \langle \ D, \ J,\ M  \ \rangle,
   \label{TreDecMass} \\
  & & {\mathfrak g}^- = \langle \ C, \ \PM{n}{\pm} \ \rangle, \quad n = \ell+ \frac{1}{2}, \cdots, 2\ell
   \nn
\eea
which satisfies the relation $ [{\mathfrak g}^0, {\mathfrak g}^{\pm} ] \subset {\mathfrak g}^{\pm} $ analogous 
to triangular decomposition. 

 Suppose the existence of the lowest weight vector $ \v0 $ defined by
\bea
  & & D \v0 = - \delta \v0, \quad \ J \v0 = -r \v0, 
   \nn \\
  & & M \v0 = -\mu \v0, \quad  X \v0 = 0, \ {}^{\forall}X \in {\mathfrak g}^-
  \label{LWVmass}
\eea
One may build a Verma module $ V^{\delta,r,\mu} $ on $ \v0 $ whose basis is given by
\beq
  \ket{k,\a,\b} = H^k \prod_{n=0}^{\ell-\hf} (\PM{n}{+})^{a_n} (\PM{n}{-})^{b_n} \v0,
  \label{basisVMmass}
\eeq
where $ \a $ and $ \b $ are $ \ell+\hf$ components vectors whose entries are non-negative integers:
\[
   \a = (a_0, a_1, \cdots, a_{\ell-\hf}), \qquad
   \b = (b_0, b_1, \cdots, b_{\ell-\hf}).
\]
Action of $ \mathfrak g $ on $ V^{\delta,r,\mu} $ is obtained by straightforward computation. 
Elements of $ {\mathfrak g}^0 $ acts on (\ref{basisVMmass}) diagonally:
\bea
   & & M \ket{k,\a,\b} = -\mu \ket{k,\a,\b}, \nn \\
   & & D \ket{k,\a,\b} = \left(-\delta + 2k + \sum_{n=0}^{\ell-1/2}2(\ell-n) (a_n + b_n) \right)
         \ket{k,\a,\b},
    \label{LA-mass-0} \\
   & & J \ket{k,\a,\b} = \left( -r + \sum_{n=0}^{\ell-1/2} (a_n - b_n) \right) \ket{k,\a,\b}.
     \nn
\eea
To have a simple expression of the formulas for $ {\mathfrak g}^{\pm} $ we introduce 
a $ \ell+1/2$ components vector $\de{i} $ having $1$ on $i$-th entry and $0$ for all 
other entries:
\[
  \de{i} = (0, \cdots, 1, \cdots, 0).
\] 
With this  $\de{i} $ action of $ {\mathfrak g}^+ $ is given by
\bea
 & & H \ket{k,\a,\b} = \ket{k+1,\a,\b}, 
  \nn \\
 & & \PM{n}{+} \ket{k,\a,\b} = \sum_{i=0}^{k} i! \binom{k}{i} 
     \binom{n}{i} \ket{k-i,\a+\de{n-i},\b},
  \label{LA-mass-+} \\
 & & \PM{n}{-} \ket{k,\a,\b} = \sum_{i=0}^{k} i! \binom{k}{i}
     \binom{n}{i} \ket{k-i,\a,\b+\de{n-i}} .
  \nn
\eea
and action of $ {\mathfrak g}^- $ is given by
\bea
 & & C \ket{k,\a,\b} = k \left( -\delta + k-1 + \sum_{n=0}^{\ell-1/2} 2(\ell-n) (a_n+b_n) \right) 
       \ket{k-1,\a,\b} 
   \nn \\
 & & \qquad 
    - \mu \left(\ell+\frac{1}{2}\right) a_{\ell-1/2} b_{\ell-1/2} I_{\ell+1/2} 
      \ket{k,\a-\de{\ell-1/2},\b-\de{\ell-1/2}} 
   \nn \\
  & & \qquad 
    + \sum_{n=0}^{\ell-3/2} (2\ell-n) \left( a_n \ket{k,\a-\de{n}+\de{n+1},\b} 
      +  b_n \ket{k,\a,\b-\de{n}+\de{n+1}} \right). 
   \nn \\
  & & \PM{n}{+} \ket{k,\a,\b} = 
     -\mu \sum_{i=0}^{n-\ell-1/2} i! \binom{k}{i}
     \binom{n}{i} b_{2\ell-n+i} I_{n-i} 
     \ket{ k-i, \a, \b-\de{2\ell-n+i}},
     \nn \\
  & & \qquad 
     + \sum_{i=n-\ell+1/2}^k i! \binom{k}{i} \binom{n}{i}
     \ket{k-i,\a+\de{n-i},\b},
     \label{LA-mas--} \\
  & & \PM{n}{-} \ket{k,\a,\b} = 
     -\mu \sum_{i=0}^{n-\ell-1/2} i! \binom{k}{i} \binom{n}{i}
     a_{2\ell-n+i} I_{n-i} 
     \ket{ k-i, \a-\de{2\ell-n+i},\b},
     \nn \\
  & & \qquad 
     + \sum_{i=n-\ell+1/2}^k i! \binom{k}{i} \binom{n}{i}
     \ket{k-i,\a, \b+\de{n-i}},
     \nn
\eea
In the above formulas we understand $ \ket{k,\a,\b} = 0 $ for $ k < 0. $ 
It is also verified by straightforward but lengthy computation that 
the above formulas are consistent with the defining commutation relations of $ \mathfrak g.$ 

  The Verma module $ V^{\delta,r,\mu} $ has a singular vector 
if $ \delta-q + (\ell+\hf)^2+1 = 0 $ for a non-negative integer $q.$ 
This is confirmed by checking that the vector 
\beq
 \ket{v_q} = ( a_{\ell} \mu H + \PM{\ell-\hf}{+} \PM{\ell-\hf}{-} )^{q} \v0, \quad 
  \alpha_{\ell} = \left( \Big(\ell-\hf \Big)! \right)^2
  \label{SVmass}
\eeq
satisfies the relations
\bea
 & & D \ket{v_q} = (2q-\delta) \ket{v_q}, \quad J \ket{v_q} = -r \ket{v_q}, \quad 
     M \ket{v_q} = -\mu \ket{v_q},
 \label{SVmassProp} \\
 & & X \ket{v_q} = 0, \quad {}^{\forall}X \in {\mathfrak g}^- \nn
\eea

%
\subsection{Differential equations with kinematical symmetry}

 Consider the space of right covariant functions (\ref{RightCovariance}) on the group 
$ G $ generated by $ {\mathfrak g} = \hg{\ell}{2}. $ 
As in the case of $ \hg{\ell}{1} $ a function $ f(g) $ in the space exhibits the property of 
the lowest weight vector of the Verma module $ V^{\delta,r,\mu} $ under the right action 
defined by (\ref{rightaction-def}). 
Parameterizing an element $ g \in \exp({\mathfrak g}^+) $ as 
\beq
  g = \exp(tH) \exp\left( \sum_{n=0}^{\ell-\hf} (x_n \PM{n}{+}+y_n\PM{n}{-})  \right),
  \label{Parametrization2mass}
\eeq
then the right action of elements in $ {\mathfrak g}^+ $ on  $ f(g) $ becomes differential operators:
\bea
  & &
    \pi_R(H) = \del{t} + \sum_{n=1}^{\ell-\hf} n \left( x_n \del{x_{n-1}} + y_n \del{y_{n-1}} \right),
  \nn \\
  & & \pi_R(\PM{n}{+}) = \del{x_n}, \qquad \pi_R(\PM{n}{-}) = \del{y_n}.
  \label{Raction2mass}
\eea
Same argument as $ \hg{\ell}{1} $ concludes that the following equation has the group $G$ generated by $ \hg{\ell}{2} $ 
as kinematical symmetry:
\bea
& &
 \left[ 
   a_{\ell} \mu \left( \del{t}  + \sum_{n=1}^{\ell-\hf} n \left( x_n \del{x_{n-1}} + y_n \del{y_{n-1}} \right) \right)
   + \frac{\partial^2}{\partial x_{\ell-\hf} \partial y_{\ell-\hf} }
 \right]^q \psi(t,x_i,y_i) = 0, 
\nn \\
& & 
 \label{InvEq2mass}
\eea
with $ q \in {\mathbb N}. $ 
For $\ell = 1/2$  (\ref{InvEq2mass}) yields the following form:
\[
  \left( \mu \del{t} + \frac{\partial^2}{\partial x \partial y} \right)^q \psi(t,x,y) = 0.
\]
By an appropriate change of variables, this recovers a hierarchy of heat/Schr\"odinger equations 
in two space dimension obtained in \cite{Mur,ADD,ADDS}.

%
\subsection{Vector field representation of $ \hg{\ell}{2} $}

 By the parametrization (\ref{Parametrization2mass}) and the left action (\ref{leftaction-def}) 
one may write down a vector field representation of $ \hg{\ell}{2} $ on the manifold $ G/G^0G^-. $ 
It is not difficult to verify the following formula of the vector field representation: 
The generators in $ {\mathfrak g}^0 $ are represented as
\bea
  & & \pi_L(D) = \delta - 2t \del{t} - \sum_{n=0}^{\ell-\hf} 2(\ell-n) \left( x_n \del{x_n} + y_n \del{y_n} \right),
  \nn \\
  & & \pi_L(J) = r - \sum_{n=0}^{\ell-\hf} \left( x_n \del{x_n} - y_n \del{y_n} \right), 
  \qquad 
  \pi_L(M) = \mu.
  \nn 
\eea
Those in $ {\mathfrak g}^+ $ are represented as
\bea
  & & 
    \pi_L(H) = -\del{t}, 
  \nn \\
  & & 
    \pi_L(\PM{n}{+}) = -\sum_{k=0}^{n} \binom{n}{k} t^k \del{x_{n-k}},
    \quad
    \pi_L(\PM{n}{-}) = -\sum_{k=0}^{n} \binom{n}{k} t^k \del{y_{n-k}},
  \nn 
\eea
and those in $ {\mathfrak g}^- $ are given by
\bea
  & & 
  \pi_L(C) = t \pi_L(D) + t^2 \del{t} + \left(\ell+\hf\right) I_{\ell+\hf} \mu x_{\ell-\hf} y_{\ell-\hf} 
  \nn \\
  & & \qquad \quad
   - \sum_{n=0}^{\ell-3/2} (2\ell-n) \left( x_n \del{x_{n+1}} + y_n \del{y_{n+1}} \right),
  \nn \\
  & & 
  \pi_L(\PM{n}{+}) = \mu \sum_{n=2\ell-n}^{\ell-\hf} \binom{n}{2\ell-k}  
       I_{2\ell-k} t^{n-2\ell+k} y_{k} 
       - \sum_{k=0}^{\ell-\hf} \binom{n}{k} t^{n-k} \del{x_k}, 
  \nn \\
  & & 
  \pi_L(\PM{n}{-}) = \mu \sum_{n=2\ell-n}^{\ell-\hf} \binom{n}{2\ell-k} 
      I_{2\ell-k} t^{n-2\ell+k} x_{k} 
        - \sum_{k=0}^{\ell-\hf} \binom{n}{k}  t^{n-k} \del{y_k}.
 \nn  
\eea

\section{$d=2$ Conformal Galilei group with exotic central extension}
\label{SEC:2exo}

\subsection{Verma module and singular vector}

 In this section we study the case of  $ \cg. $ 
Contrast to the analysis in \S \ref{SEC:1mass} and \S \ref{SEC:2mass}, $ \ell $ takes a positive 
integral values for $ \cg. $ 
We redefine the generators to introduce the decomposition analogous to triangular one:
\beq
  \PE{n}{\pm} = \PE{n}{1} \pm i \PE{n}{2}, \qquad J = -i M_{12}, \qquad 
  \Theta \to -2i \Theta.
  \label{NewGenerators}
\eeq 
Then the nonvanishing commutators of $ \cg $ are identical to (\ref{ExoticCGAdef}) except the 
central extension. The central extension for $\cg$ in terms of new generators is given by 
\beq
 [\PE{m}{\pm}, \PE{n}{\mp} ] = \pm \delta_{m+n,2\ell} I_m \Theta
 \label{CentExtNewGenExo}
\eeq
with $ I_m $ defined in (\ref{ExoticExtension}). 
We make a triangular like decomposition of $ {\mathfrak g} \equiv \cg: $ 
\bea
  & & {\mathfrak g}^{+} = \langle \ H, \ \PE{\ell}{+}, \ \PE{n}{\pm} \ \rangle, \quad n = 0, 1, \cdots, \ell- 1 
  \nn \\
  & & {\mathfrak g}^{0} = \langle \ D, \ J, \ \Theta \ \rangle, 
  \label{TriDec} \\ 
  & & {\mathfrak g}^{-} = \langle \ C, \ \PE{\ell}{-}, \ \PE{n}{\pm} \ \rangle, \quad n = \ell+1, \ell+2, \cdots, 2\ell 
  \nn
\eea
It is an easy task to see that $ [{\mathfrak g}^0, {\mathfrak g}^{\pm}] \subset {\mathfrak g}^{\pm}. $ 

 The lowest weight vector $ \u0 $ is defined as usual:
\bea
 & & 
   D \u0 = -\delta \u0, \quad \ J \u0 = -r \u0, 
   \nn \\
 & &  \Theta \u0 = \theta \u0, \qquad X \u0 = 0 \ \mbox{for } {}^{\forall}X \in {\mathfrak g}^{-}.
   \label{LWdef}
\eea
We construct a Verma module $ V^{\delta,r,\theta} $ by repeated applications of an element of ${\mathfrak g}^+$ 
on $ \u0. $ In order to specify a basis of $ V^{\delta,r,\theta}$ we introduce the following notations. 
Let $ \a$ and $ \b $ be $(\ell+1) $ and $ \ell$ components vectors, respectively. 
Their entries are non-negative integers:
\[
 \underline{a} = (a_0, a_1, \cdots, a_{\ell}), \qquad 
 \underline{b} = (b_0, b_1, \cdots, b_{\ell-1}).
\]
We also introduce $(\ell+1)$ components vector $ \ep{i} $ with 1 in the $i$th entry and 
0 elsewhere. Similarly, $\ell$ components vector $ \de{i} $ having 1 in the $i$th entry and 
0 elsewhere, \textit{i.e.},
\[
  \ep{i} = (0, \cdots, 1, \cdots, 0), \qquad 
  \de{i} = (0, \cdots, 1, \cdots, 0)
\]
With these notations the following vectors form a basis  of $ V^{\delta,r,\theta} $
\beq
  \ket{h, \a, \b} = H^h ( \PE{\ell}{+} )^{a_{\ell}} 
  \prod_{n=0}^{\ell-1} ( \PE{n}{+} )^{a_n} (  \PE{n}{-} )^{b_n} \v0.
  \label{basisVMexotic}
\eeq
It is not difficult, but with a lengthy computation, to write down the action of $ \mathfrak g $ 
on $ V^{\delta,r,\theta} $ and to verify that $ V^{\delta,r,\theta}$ indeed carries a representation of $\mathfrak g.$ 
The action of $ {\mathfrak g}^0 $ on $ \ket{h,\a,\b} $ is diagonal:
\bea
  & & \Theta \ket{h,\a,\b} = \theta \ket{h,\a,\b},
  \nn \\
  & & D \ket{h,\a,\b} = \left(-\delta + 2h +\sum_{n=0}^{\ell-1} 2(\ell-n) (a_n + b_n) \right) \ket{h,\a,\b},
  \label{action-g0} \\
  & & J \ket{h,\a,\b} \left( -r  + \sum_{n=0}^{\ell-1}(a_n-b_n) + a_{\ell} \right) \ket{h,\a,\b}.
  \nn
\eea
While the action of $ {\mathfrak g}^+ $ is given by
\bea
 & & H \ket{h,\a,\b} =  \ket{h+1,\a,\b},
 \nn \\
 & & \PE{n}{+}  \ket{h,\a,\b} = 
     \sum_{i=0}^h i! \binom{h}{i} \binom{n}{i}  \ket{h-i,\a+\ep{n-i},\b},
 \label{action-g+} \\
 & & \PE{n}{-}  \ket{h,\a,\b} = 
     \sum_{i=0}^h i! \binom{h}{i} \binom{n}{i} \ket{h-i,\a,\b+\de{n-i}},
  \nn
\eea
and the action of $ {\mathfrak g}^- $ by
\bea
 & & 
  C \ket{h,\a,\b} 
  = h \left( -\delta+h-1 + \sum_{n=0}^{h-1} 2(\ell-n)(a_n+b_n) \right) \ket{h-1,\a,\b}
  \nn \\
 & & \qquad
   + \; \ell a_{\ell} b_{\ell-1} I_{\ell+1} \theta \ket{h,\a-\ep{\ell},\b-\de{\ell-1}}
  \nn \\
 & &\qquad
   + \sum_{n=0}^{\ell-1} (2\ell-n) a_n \ket{h,\a-\ep{n}+\ep{n+1},\b}
 \nn \\
 & & \qquad 
   + \sum_{n=0}^{\ell-2} (2\ell-n) b_n \ket{h,\a,\b-\de{n}+\de{n+1}},
  \nn \\
 & & \PE{n}{+} \ket{h,\a,\b} 
  = \sum_{i=0}^{n-\ell-1} i! \binom{h}{i} \binom{n}{i}  
   b_{2\ell-n+i} I_{n-i} \theta \ket{h-i,\a,\b-\de{2\ell-n+i}}
  \nn \\
 & & \qquad
  + \sum_{i=n-\ell}^{h} i! \binom{h}{i} \binom{n}{i}  \ket{h-i,\a+\ep{n-i},\b}, 
   \label{action-g-} \\
 & & \PE{n}{-} \ket{h,\a,\b} 
  = -\sum_{i=0}^{n-\ell} i! \binom{h}{i} \binom{n}{i}  
   a_{2\ell-n+i} I_{n-i} \theta \ket{h-i,\a-\ep{2\ell-n+i},\b}
  \nn \\
 & & \qquad
  + \sum_{i=n-\ell+1}^{h} i! \binom{h}{i} \binom{n}{i} \ket{h-i,\a,\b+\de{n-i}}, 
   \nn
\eea
In these formulas we understand $ \ket{h,\a,\b} = 0 $ for $ h < 0. $

  The Verma module $ V^{\delta,r,\theta} $ has a singular vector if 
$ \delta -q + \ell(\ell+1) + 1 = 0 $ for a positive integer $q$ given by
\beq
 \ket{v_q} = ( \alpha_{\ell} \theta H + (-1)^\ell \PE{\ell-1}{-} \PE{\ell}{+} )^{q} \u0,
 \quad 
 \alpha_{\ell} =  \ell! (\ell-1)!
 \label{SVexotic}
\eeq
It is straightforward to verify that the vector (\ref{SVexotic}) satisfies
\bea
 & & 
  D \ket{v_q} = (2q-\delta) \ket{v_q}, \quad J \ket{v_q} = -r \ket{v_q}, \quad \Theta \ket{v_q} = \theta \ket{v_q},
  \nn \\
 & & 
  X \ket{v_q} = 0, \quad {}^{\forall}X \in {\mathfrak g}^-
  \label{SVexoticProp}
\eea
This generalizes the result for $ \ell = 1 $ investigated in \cite{NAPSI}. 

\subsection{Differential equations with kinematical symmetry}

 We parametrize an element of $ g \in \exp({\mathfrak g}^+) $ as 
\beq
 g = e^{tH} \exp\left(  \sum_{n=0}^{\ell-1} ( x_n \PE{n}{+} + y_n \PE{n}{-}) + x_{\ell} \PE{\ell}{+} \right).
 \label{Parametrization2exotic}
\eeq
A right covariant function on $ G = exp(\cg) $ exhibits the properties of the lowest weight vector (\ref{LWdef}). 
It is also easily seen that the right action (\ref{rightaction-def}) of $ {\mathfrak g}^+ $ is given by
\bea
& &
 \pi_R(H) = \frac{\partial}{\partial t} + \sum_{n=0}^{\ell-1} (n+1) x_{n+1} \frac{\partial}{\partial x_n} 
          + \sum_{n=0}^{\ell-2} (n+1) y_{n+1} \frac{\partial}{\partial y_n},
 \nn \\
& &
 \pi_R( \PE{n}{+} ) = \frac{\partial}{\partial x_n}, \quad 
 \qquad
 \pi_R( \PE{n}{-} ) = \frac{\partial}{\partial y_n}.
 \label{Raction2exotic}
\eea
It follows together with (\ref{SVexotic}) that following equations have the group generated by $ \cg $ as kinematical symmetry:
\bea
 & &
  \left[ 
     \alpha_{\ell} \theta 
       \left(
          \frac{\partial}{\partial t} + \sum_{n=1}^{\ell} n x_{n} \frac{\partial}{\partial x_{n-1}} 
          + \sum_{n=1}^{\ell-1} n y_{n} \frac{\partial}{\partial y_{n-1}}
       \right)
       + (-1)^{\ell} \frac{\partial^2}{\partial y_{\ell-1} \partial x_{\ell}} 
  \right]^q \psi = 0.
 \nn \\
 & & 
  \label{InvEq2exotic}
\eea
One may see the symmetry by the same agrument as \S \ref{EqSymm}.

%
\subsection{Vector field representation of $ \cg $}

 The parametrization (\ref{Parametrization2exotic}) and the left action (\ref{leftaction-def}) 
gives the following vector field representation of $ \cg $ on the coset $ G/G^0G^-. $ 
Generators in $ {\mathfrak g}^0 $ are represented as
\bea
  & & \pi_L(D) = \delta -2t \der{t}
      - \sum_{n=0}^{\ell-1} 2(\ell-n) \left( x_n \der{x_n} + y_n \der{y_n} \right),
  \nn   \\
  & & \pi_L(J) = r - \sum_{n=0}^{\ell} x_n \der{x_n} + \sum_{n=0}^{\ell-1} y_n \der{y_n},
  \qquad 
     \pi_L(\Theta) = - \theta.
 \nn
\eea
While the generators in $ {\mathfrak g}^+ $ are represented as 
\bea
& & 
  \pi_L(H) = - \der{t}, \qquad
  \pi_L(\PE{n}{+}) = -\sum_{k=0}^n \binom{n}{k} t^k \der{x_{n-k}},
 \nn \\
& & 
  \pi_L(\PE{n}{-}) =  -\sum_{k=0}^n \binom{n}{k} t^k \der{y_{n-k}}.
  \nn
\eea
and those in $ {\mathfrak g}^- $ as 
\bea
 & &
   \pi_L(C) =  t \pi_L(D) + t^2\del{t}  -\ell I_{\ell+1}\theta x_{\ell} y_{\ell-1} 
   - \sum_{n=0}^{\ell-1}(2\ell-n)x_n \del{x_{n+1}} 
  \nn \\
 & & \qquad   \ 
  - \sum_{n=0}^{\ell-2} (2\ell-n) y_n \del{y_{n+1}},
  \nn \\
 & & \pi_L(\PE{n}{+}) = - \theta \sum_{k=0}^{n-\ell+1} \binom{n}{k} I_{n-k} t^k \, y_{2\ell-n+k} 
  - \sum_{k=n-\ell}^{n} \binom{n}{k} t^k \der{x_{n-k}},
  \nn \\
 & & \pi_L(\PE{n}{-}) = \theta \sum_{k=0}^{n-\ell} \binom{n}{k} I_{n-k} t^k \, x_{2\ell-n+k}
  - \sum_{k=n-\ell+1}^{n} \binom{n}{k} t^k \der{y_{n-k}}.
  \nn
\eea

\section{Concluding remarks}
\label{SEC:CR}

 We have investigated the representations of the $ \ell$-conformal Galilei algebra with 
central extensions. For $ d = 2 $ we showed that some Verma modules are not irreducible by 
giving the singular vectors explicitly. By employing the method in \cite{Dob} we derived 
partial differential equations having the group generated by $ \hg{\ell}{1}$ or $  \hg{\ell}{2} $ or 
$ \cg $ as kinematical symmetry. 
Obtained differential equations have some common properties. They form a hierarchy of 
linear differential equations. 
Each hierarchy contains precisely one differential equation of second order. 
This is a big difference of the present result from the previous works considering 
physical systems relating to $\ell$-conformal Galilei group with higher $\ell.$ 

 We restrict ourselves to the algebras of $ d = 1, 2 $ and having central extension. 
One may repeat the same analysis for higher values of $ d $ or the algebras without 
central extension. Representation theory may be more involved for such cases. However, 
the case of $ d = 3 $ has a special interest since spatial rotations become non-Abelian and 
have a contribution to all sectors in triangular decomposition. This implies that  
differential equation will have different structure from the cases of $ d = 1, 2. $ 
Another important, but mathematical, problem to be done is a precise investigation of 
irreducible representations. Namely, one may try to classify irreducible Verma modules 
for $ d > 2 $ and higher $ \ell $ as done in \cite{DoDoMr,AIK,NAPSI}. 
These will be a future work.

\ack

N.A. is supported by a grants-in-aid from JSPS (Contract No.23540154).

\appendix
\section*{Appendix}
\setcounter{section}{1}

  In this Appendix we investigate the centerless algebra $ \g{1}{1}, $ i.e., $ (d,\ell) = (1,1). $ 
This is motivated by the recent studies of the field theory potentially dual to 3d Minkowski spacetime  \cite{Bagchi,BaFare,BaGoGo}. 
The dimension of the algebra $ \g{1}{1}$ is six  and the generators are denoted by  $ D, H, C, \Pone{0}, \Pone{1}, \Pone{2}. $ 
Nonvanishing commutators are given by
\begin{eqnarray}
  & & [D, H] = 2H, \qquad [D, C]=-2C, \qquad [C, H]=D,  
  \nonumber \\
  & & [H, \Pone{n}] = -n \Pone{n-1}, \quad 
      [D, \Pone{n}] = 2(1-n) \Pone{n}, \nonumber \\
  & & 
      [C, \Pone{n}] = (2-n) \Pone{n+1}. 
      \label{BosonicDef11} 
\end{eqnarray}
We introduce the triangular-like decomposition:
\[
  \g{1}{1}^-  = \{ \; H, \ \Pone{0} \; \}, \quad 
  \g{1}{1}^0 = \{ \; D, \ \Pone{1} \; \}, \quad 
  \g{1}{1}^+ = \{ \; C, \ \Pone{2} \; \}, \quad 
\]
The lowest weight vector $ \ket{\delta,\kappa} $ and the Verma module $ V^{\delta,\kappa} $ are defined as usual:
\bea
 & & 
  D \ket{\delta,\kappa} = -\delta \ket{\delta,\kappa}, \quad \Pone{0} \ket{\delta,\kappa} = -\kappa \ket{\delta,\kappa},
  \nn \\
 & & 
  H \ket{\delta,\kappa} = \Pone{0} \ket{\delta,\kappa} = 0. 
  \label{LWV11}
\eea
\beq
  V^{\delta,\kappa} = \{ \; C^h (\Pone{2})^k \ket{\delta,\kappa} \ | \ h, k \in {\mathbb Z}_{\geq 0} \ \}.
  \label{VM11}
\eeq
Writing $ \ket{h,k} = C^h (\Pone{2})^k \ket{\delta,\kappa} $ we have
\beq
  D \ket{h,k} = -(\delta+2h+2k) \ket{h,k}. \label{DonVM}
\eeq
We set $ p = h+k $ and call the nonnegative integer $p$ \textit{level}. 
Then the following statements on the existence of singular vectors in $ V^{\delta, \kappa} $ hold true. 
\begin{itemize}
  \item if $ \kappa \neq 0 $ then there exists no singular vector in $ V^{\delta,\kappa}.$ 
  \item if $ \kappa = 0 $ there exist unique singular vector at each level $p$ and it is given (up to an overall constant) by
  $ (\Pone{2})^p \ket{\delta,0}.$ 
\end{itemize}
Thus $ V^{\delta, \kappa} $ is irreducible if $ \kappa \neq 0. $ 
One may prove them in a way similar to \cite{DoDoMr,NAPSI}, although we omit the proof. 
Parametrizing the group element as $ g = \exp(tH + x\Pone{2}) $ the right action of $ \Pone{2} $ is given by
\beq
 \pi_R(\Pone{2}) = \der{x}. \label{RA11}
\eeq
This leads the equations having the group generated by $ \g{1}{1} $ as kinematical symmetray:
\beq
  \left( \der{x} \right)^p \psi(t,x) = 0. \label{IE11}
\eeq
The equations are not of physical interest. This is the situation similar to the Schr\"odinger algebra without 
central extension \cite{DoDoMr}.

\end{document}